\documentclass[12pt]{article}
\usepackage{amsfonts,latexsym,eucal,color,graphicx}
\usepackage{epsfig,amssymb,amsmath,cite}
\usepackage{graphicx}
\definecolor{red  }{rgb}{1,0,0}
\definecolor{blue }{rgb}{0,0,1}
\definecolor{green}{rgb}{0,1,0}
\newcommand{\dalm}{\kern1pt\vbox{\hrule height 0.9pt\hbox{\vrule width 0.9pt\hskip 2.5pt\vbox{\vskip 5.5pt}\hskip 3pt\vrule width 0.3pt}\hrule height 0.3pt}\kern1pt}

\begin{document}
% \draft command makes pacs numbers print 
% \draft

\begin{titlepage}

\begin{flushright}
{WU-AP/261/07}
%%{gr-qc/yymmnnn}
\end{flushright}
\vspace{1cm}

%---------------------------------------------------------------------%
\begin{center}
	{\Large
		{\bf 
			On Hawking radiation from black rings
		}
	}
\end{center}
\vspace{1cm}
%---------------------------------------------------------------------%

\begin{center}
Umpei Miyamoto$^{a,1}$ and Keiju Murata$^{b,2}$

%%--------   Address  ------------------
\vspace{.5cm}
{\small {\textit{$^{a}$
Department of Physics,
 Waseda University, Okubo 3-4-1, Tokyo 169-8555, Japan, 
}}
}
\\
\vspace{5mm}
{\small \textit{$^{b}$
Department of Physics, Kyoto University, Kyoto 606-8501, Japan
}}
\\
%%----------------------------------------
\vspace*{1.0cm}

%%----------- Email  ----------------------
{\small
{\tt{
\noindent
$^{1}$ umpei\;\!@gravity.phys.waseda.ac.jp
\\
%\noindent
$^{2}$ murata\;\!@tap.scphys.kyoto.ac.jp
}}
}
%%----------------------------------------
\end{center}

\vspace*{1.0cm}

%---------------------------------------------------------------------%
%---------------------------------------------------------------------%

%\date{\today}
%\preprint{2006-09-01, WU-AP/254/06, UTAP-***, gr-qc/06*****}
%---------------------------------------------------------------------%

\begin{abstract}
We calculate the quantum radiation from the five dimensional charged
 rotating black rings
by demanding the radiation to eliminate the possible anomalies on the
 horizons.
It is shown that the
 temperature, energy flux and angular-momentum flux exactly coincide
 with those of the Hawking radiation.
The black rings considered in this paper contain the Myers-Perry black
 hole as a limit and the quantum radiation for this black hole, obtained
 in the literature, is recovered in the limit.
The results support the picture that the Hawking radiation can be
 regarded as the anomaly eliminator on horizons and suggest its general
 applicability to the higher-dimensional black holes discovered
 recently.
\end{abstract}

\end{titlepage}

%%%-----------------------------------------------------------------%%%
%%%-----------------------------------------------------------------%%%
\section{ Introduction }
%%%-----------------------------------------------------------------%%%
%%%-----------------------------------------------------------------%%%

One of the outstanding predictions by the quantum field theory in curved spacetime is the evaporation of black holes, shown by Hawking~\cite{Hawking:1974sw}.
The particles radiated from black holes are characterized by the thermal spectrum with the temperature $ T = (1/ 2 \pi) \kappa $, where $\kappa$ is the surface gravity of the black hole. This fixes the coefficient between the entropy and horizon area, $ S = ( 1/ 4) A $, and results in the intensive studies on the microscopic origin of the entropy.

To give black holes the degree of freedom to possess the entropy, it is believed that the quantization of spacetimes is needed. It will be helpful, however, to interpret the thermal radiation from the viewpoint of the semiclassical revel at this time. Recently, an interesting interpretation was proposed that the Hawking radiation is a energy flow compensating the possible gravitational anomaly resulting from the chiral nature of the effective theory near horizons~\cite{Robinson:2005pd}. In fact, the Hawking temperature of spherical black holes was reproduced from this viewpoint~\cite{Robinson:2005pd}. In Ref.~\cite{Iso:2006wa}, it was shown that the gauge anomaly also has to be taken into account to obtain the correct charge flow radiated by a charged black hole. A further interesting observation is that the rotation of black holes appears as a $U(1)$-gauge field in the effective theory and, in fact, the angular-momentum flow was calculated correctly by demanding the cancellation of the ``gauge'' anomaly~\cite{Iso:2006ut,Iso:2006xj}. It should be noted that the correct temperature and currents of (singly) rotating black hole can be estimated only via the cancellation of the gravitational anomaly~\cite{Murata:2006pt}. See also Refs.~\cite{Xu:2006tq,Setare:2006hq,Vagenas:2006qb,Kui:2007dy,Jiang:2007wj,Jiang:2007pn,Jiang:2007gc,Shin:2007gz} for the applications for various black holes and~\cite{Das:2007ru} for a short review.

The idea to relate the Hawking radiation with anomalies traces back to the seminal work by Christensen and Fulling~\cite{Christensen:1977jc}, in which they showed that the Hawking radiation can be regarded as a conformal anomaly. There is a crucial point in this interpretation, however, that the interpretation is effective only for the systems possessing the conformal invariance. In particular, the correct prediction (i.e., the flux of Hawking radiation) is possible only in two dimensions. The idea in \cite{Robinson:2005pd} has the opposite direction in the sense that the Hawking radiation plays the role to cancel the quantum anomalies, rather than the Hawking radiation itself is regarded as an anomaly. 
The crucial point in the case of black holes is that a quantum field, e.g., a massless scalar field, near the horizons can be reduced to the system of an infinite number of fields in ($1+1$)-dimensional spacetime. Since the ingoing modes on the horizon cannot affect the physics outside the horizon classically, if we ignore them, the effective theory becomes chiral and the diffeomorphism invariance is violated at quantum revel.
For spherically symmetric black holes, the reduction to the effective theory in the ($t,r$)-sector ($r$ is a radial coordinate) seems to be trivial relatively, and the procedure is known to be applicable for for rotating black holes~\cite{Iso:2006xj,Murata:2006pt} at least.
In this paper, we consider 5-dimensional rotating black rings, of which horizon topology is $ S^1 \times S^2 $. One might think that for these solutions, the effective theory could not be reduced to two dimensional one. We will show the reduction procedure is possible despite of the non-triviality of the horizon topology. Then, the temperature and the fluxes of angular momentum and total energy are calculated explicitly with the condition of anomaly cancellation, and shown to coincide exactly with ones calculated from a Planckian distribution function. In other words, we show that the Hawking radiation of black rings are capable of eliminating the possible anomalies on the horizons.

The black rings which we consider in this paper is the so-called dipole black rings~\cite{Emparan:2004wy,Yazadjiev:2006hw}, which contain the Emparan-Reall black ring~\cite{Emparan:2001wn} and also contain the Myers-Perry black hole~\cite{Myers:1986un} in suitable limits. This solution is used to show an infinite non-uniqueness of black ring solutions since it does not have a conserved net charge but local distribution of charge. That is, there are infinitely many solutions for fixed mass and angular-momentum.
Although the thermodynamic properties (and therefore the Hawking radiation) of higher-dimensional black holes~\cite{Nozawa:2005eu,Arcioni:2004ww} are important to picture their phase structure and evolutions~\cite{Elvang:2007hg}, the Hawking radiation of black rings has not been investigated sufficiently so far. This situation seems to originate from the difficulty to separate variables in ring spacetimes.
The point should be stressed that our technique does not need the separation of variables and explicit harmonic functions since the properties of the Hawking radiation can be determined only by the near horizon physics.

The organization of this paper is as follows.
In Sec.~\ref{sec:preparation}, we introduce the dipole black rings and
their properties necessary for the following analysis. 
We also calculate the energy and angular-momentum fluxes of the Hawking
radiation from a Planckian spectrum. In
Sec.~\ref{sec:radiation}, 
the behavior of a quantum field near the horizons is investigated and 
we estimate the quantum fluxes of angular
momentum and energy radiated from the black rings by demanding the anomaly cancellation.
The limit to the Myers-Perry black hole
is also discussed there. The final section is devoted to a conclusion. We
use the units in which $ c = G = \hbar = k_B = 1  $ and the almost plus
notation of the metric throughout this paper.

%\vspace{0.2cm}
%Gravitational anomaly~\cite{Bertlmann:2000da}.
%Bardeen-Zumino current~\cite{Bardeen:1984pm}.

%%%-----------------------------------------------------------------%%%
%%%-----------------------------------------------------------------%%%
\section{ Dipole black rings and Hawking radiation}
\label{sec:preparation}
%%%-----------------------------------------------------------------%%%
%%%-----------------------------------------------------------------%%%

In this section, we introduce the dipole black ring solution and its
properties needed for the following analysis. We also derive the 
Hawking fluxes of the dipole rings by integrating the thermal spectrum. 
%Then, we show that
%scalar field near the horizon of the ring can be regarded as two
%dimensional field in the presence of a gauge field, which is crucial
%for the calculation of the Hawking radiation via anomaly cancelation. 

%%%-----------------------------------------------------------------%%%
\subsection{ Dipole black rings }
\label{subsec:dipole-ring}
%%%-----------------------------------------------------------------%%%

%The analytic solution of rotating regular black ring in 5-dimensional general relativity was first obtained by Emparan and Reall in~\cite{Emparan:2001wn}.
%Then, supersymmetric/non-supersymmetric black rings were obtained and their classical and microscopic properties were extensively studied (see~\cite{Emparan:2006mm} for a review). Here, we consider the so-called dipole black ring solutions~\cite{Emparan:2004wy}, which has a local charge distribution and was discussed in the context of infinite non-uniqueness of the black rings.

\begin{figure}[hb]
\begin{center}
\includegraphics[width=7cm]{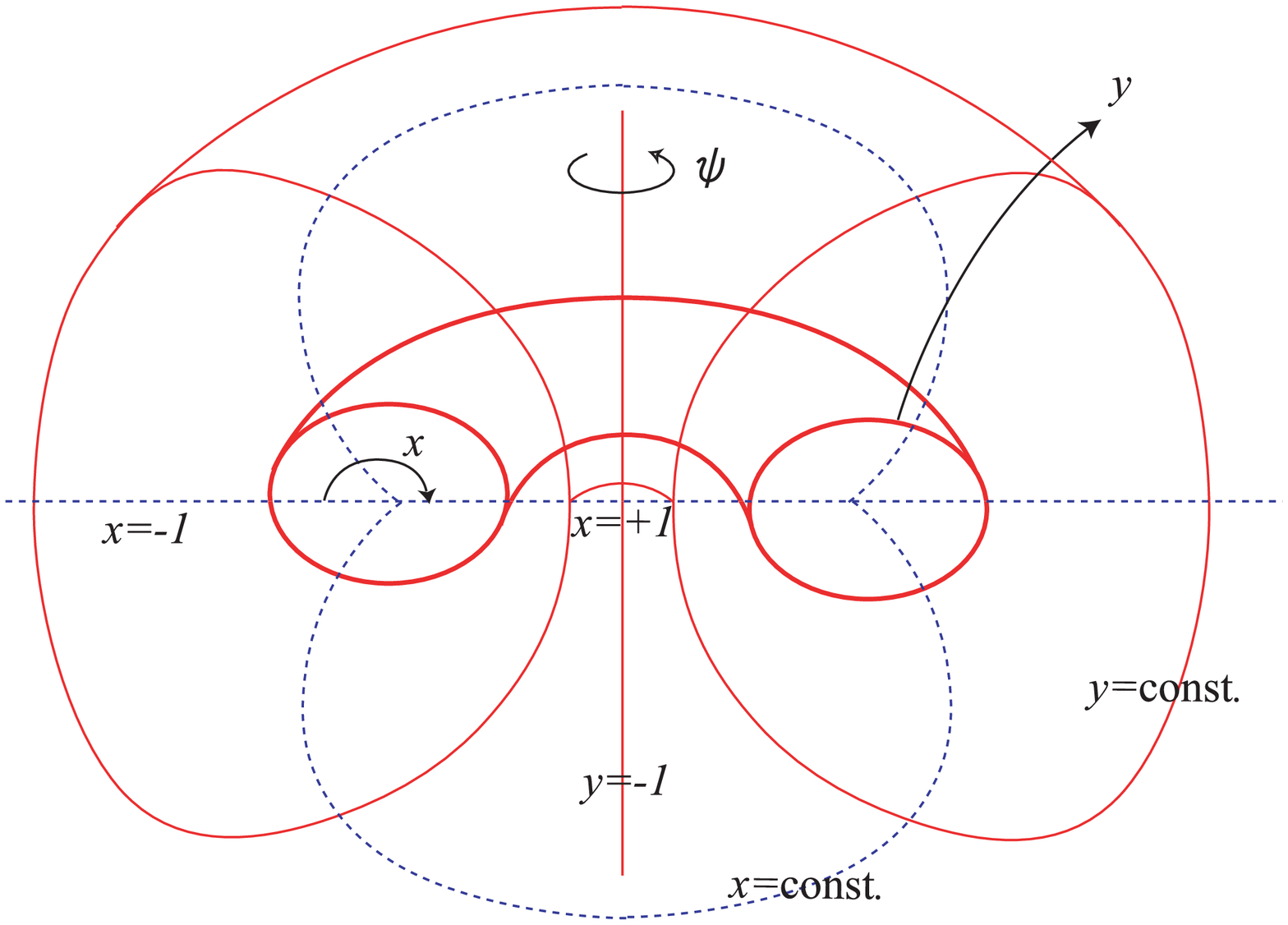}
\caption{
A schematic picture of the ring coordinates for the metric Eq.~(\ref{eq:ring})
at constant $t$ and $\phi$. The red (solid) surfaces and blue (dotted) curves represent constant $y$ and $x$, respectively. In particular, the bold red (solid) surface is the event horizon, given by $y = y_h := -1/\nu $. Infinity lies at $ x=y=-1 $. This black ring solution approaches the Myers-Perry black hole, of which topology is $S^3$, in the particular limit described in Sec.~\ref{subsec:dipole-ring}.
\label{fg:ring-coord}}
\end{center}
\end{figure}

Let us consider the following 5-dimensional action, which is obtained by dualizing the Einstein-Maxwell-dilaton system~\cite{Emparan:2004wy}:
\begin{eqnarray}
	I 
	=
	\frac{ 1 }{ 16\pi }
	\int
	d^5 \! x \sqrt{-g}
	\left[
		\mathcal{R} - \frac{ 1 }{ 2 } (\partial \Phi)^2 - \frac{ 1 }{ 12 } e^{ -\alpha \Phi } \mathcal{H}^2
	\right],
\end{eqnarray}
where $ \mathcal{H} $ is a three-form field strength and $ \Phi $ is a dilaton.
The dipole black ring solution in this system, which can represent either a magnetic or an electric black rings, is given by
\begin{eqnarray}
	&&
	ds^2
	=
	-
	\frac{ F(y) }{ F(x) }
	\left[
		\frac{ H(x) }{ H(y) }
	\right]^{ N/3 }
	\left[
		dt - C R \frac{ 1+y }{ F(y) } d\psi
	\right]^2
	+
	\frac{ R^2 }{ (x-y)^2 } F(x)
	\left[
		H(x) H^2(y)
	\right]^{ N/3 }
	\times
	\nonumber
	\\
	&&
	\hspace{4cm}
	\times
	\left[
		- \frac{ G(y) }{ F(y) H^N(y) } d\psi^2
		- \frac{ dy^2 }{ G(y) }
		+ \frac{ dx^2 }{ G(x) }
		+ \frac{ G(x) }{ F(x) H^N(x) } d\phi^2
	\right],
	\label{eq:ring}
\end{eqnarray}
where
\begin{eqnarray}
	&&
	F(s)
	:=
	1 + \lambda s,
	\;\;\;\;\;
	G(s)
	:=
	( 1-s^2 ) ( 1+\nu s ),
	\;\;\;\;\;
	H(s)
	:=
	1-\mu s,
\end{eqnarray}
and
\begin{eqnarray}
	C
	:=
	\sqrt{
		\lambda ( \lambda - \nu ) \frac{  1+\lambda }{ 1-\lambda }
	}.
\end{eqnarray}
The dimensionless constants $\nu$, $\lambda$ and $ \mu $ lie in the range
\begin{eqnarray}
	0
	<
	\nu
	\leq
	\lambda
	<
	1,
	\;\;\;\;\;
	0
	\leq
	\mu
	<1.
\end{eqnarray}
The constant $ R $ has the dimension of length and for thin large rings corresponds roughly to the radius of the ring circle~\cite{Elvang:2003mj}. The dimensionless constant $ N $ is related to the dilaton coupling by
\begin{eqnarray}
	\alpha^2
	=
	\frac{ 4 }{ N }
	-
	\frac{ 4 }{ 3 },
	\;\;\;\;\;
	0
	<
	N
	\leq
	3.
\end{eqnarray}
It is noted that the values $ N=1,2,3 $ are of particular relevance to string and M-theory~\cite{Emparan:2004wy}.
Taking the limit of $ \mu \to 0 $ in Eq.~(\ref{eq:ring}), we recover the neutral black ring found in~\cite{Emparan:2001wn}.
The coordinates $ x $ and $ y $ vary within the ranges
\begin{eqnarray}
	-1
	\leq
	x
	\leq
	1,
	\;\;\;\;\;
	- \infty
	\leq
	y
	\leq
	-1.
	\label{eq:range}
\end{eqnarray}
See Fig.~\ref{fg:ring-coord} for the visualization of this black ring and its coordinates.
The possible conical singularities at the axes extending to infinity, $ x=-1 $ and $ y=-1 $, are avoided by setting the periods of the angular coordinates as
\begin{eqnarray}
	\Delta \psi
	=
	\Delta \phi
	=
	4\pi
	\frac{ H^{ N/2 }(-1) \sqrt{ F(-1) } }{ | G^\prime (-1) | }
	=
	2\pi
	\frac{ ( 1+\mu )^{ N/2 } \sqrt{ 1-\lambda } }{ 1-\nu }.
	\label{eq:deficit}
\end{eqnarray}
With one more additional condition to avoid the conical singularity at
$ x=+1 $~\cite{Emparan:2004wy}, it is shown that the solution has a
regular even horizon at $ y = y_h := -1/\nu $.\footnote{In addition,
there is an inner horizon at $ y = - \infty $. The metric can be
continued beyond this horizon to $ 1/\mu < y < \infty $. The two
horizons coincide when $ \nu=0 $, and therefore $ \nu $ is regarded as a
non-extremality parameter. We do not describe this point further since
it is sufficient for us to consider the outer region of the event
horizon in this paper.}

%One can calculate the mass, angular momentum, horizon area, temperature
%and angular velocity at the
%outer horizon as 
%\begin{align}
%  M &= \frac{3\pi R^2}{4G}
%  \frac{(1+\mu)^N}{1-\nu}\left(\lambda+\frac{N}{3}\frac{\mu(1-\lambda)}{1+\mu}\right)\
%  ,\\
%  J &= \frac{\pi
%  R^3}{2G}\frac{(1+\mu)^{3N/2}\sqrt{\lambda(\lambda-\nu)(1+\lambda)}}{(1-\nu)^2}\
%  ,\\
%  \mathcal{A}_H &= 8\pi^2 R^3\frac{(1+\mu)^N
%  \nu^{(3-N)/2}(\mu+\nu)^{N/2}\sqrt{\lambda(1-\lambda^2)}}{(1-\nu)^2(1+\nu)}\
%  ,\\
%  T &=
%  \frac{1}{4\pi R}\frac{\nu^{(N-1)/2}(1+\nu)}{(\mu+\nu)^{N/2}}\sqrt{\frac{1-\lambda}{\lambda(1+\lambda)}}\
%  ,\label{eq:T_H}\\
%  \Omega &=
%  \frac{1}{R}\frac{1}{(1+\mu)^{N/2}}\sqrt{\frac{\lambda-\nu}{\lambda(1+\lambda)}}\ \label{eq:Omega_H}.
%\end{align}

Here, let us see the neutral black ring, Eq.~(\ref{eq:ring}) with $\mu=0$, describes the Myers-Perry black hole in a particular limit~\cite{Myers:1986un,Harmark:2004rm}. Before taking the limit, we introduce the new parameters ($ M, a $) and coordinates ($r,\theta$) given by
\begin{eqnarray}
	M
	:=
	\frac{ 2R^2 }{ 1-\nu },
	\;\;\;\;\;
	a^2
	:=
	2R^2
	\frac{ \lambda-\nu }{ (1-\nu)^2 },
\end{eqnarray}
and 
\begin{eqnarray}
	&&
	x
	=
	-1
	+
	2
	\left(
		1- \frac{ a^2 }{ M }	
	\right)
	\frac{ R^2 \cos^2 \theta }{ r^2 - ( M-a^2 ) \cos^2 \theta },
	\nonumber
	\\
	&&
	y
	=
	-1
	-2 \left(
		1- \frac{ a^2 }{ M }	
	\right)
	\frac{ R^2 \sin^2 \theta }{ r^2 - ( M-a^2 ) \cos^2 \theta }.
\end{eqnarray}
In addition, we rescale $ (\psi,\phi) \to \sqrt{ (M-a^2)/ (2 R^2) } \; (\psi,\phi) $ so that they have canonical periodicity $ 2\pi $.
Then, taking the limit in which $ \lambda,\nu \to 1 $ and $ R \to 0 $ with $M$ and $a$ kept finite, we have
\begin{eqnarray}
	ds^2
	=
	-\left(
		1 - \frac{ M }{ \Sigma }
	\right)
	\left(
		dt - \frac{ M a \sin^2 \theta }{ \Sigma - M } d\psi
	\right)^2
	+
	\Sigma \left( \frac{ dr^2 }{ \Delta } + d\theta^2 \right)
	+
	\frac{ \Delta \sin^2 \theta }{ 1-M/\Sigma } d\psi^2
	+
	r^2 \cos^2 \theta d\phi^2,
	\nonumber
	\\
	\label{eq:MP}
\end{eqnarray}
where
\begin{eqnarray}	
	\Delta
	:=
	r^2 - M + a^2,
	\;\;\;\;\;
	\Sigma 
	:=
	r^2 + a^2 \cos^2 \theta.
\end{eqnarray}
This spacetime, Eq.~(\ref{eq:MP}), is nothing but the 5-dimensional Myers-Perry black hole with the rotation in the $ \psi $-direction, of which horizon topology is $ S^3 $.

%%%-----------------------------------------------------------------%%%
%%%-----------------------------------------------------------------%%%

%%%-----------------------------------------------------------------%%%
%%%-----------------------------------------------------------------%%%
\subsection{Hawking fluxes evaluated from the thermal spectrum }
\label{sec:fluxes}
%%%-----------------------------------------------------------------%%%
%%%-----------------------------------------------------------------%%%

We will calculate the angular-momentum and energy fluxes
of the Hawking radiation from the viewpoint of anomaly cancellation later.
Before such a calculation, we derive the fluxes by integrating the thermal spectrum.

The thermal distribution of the Hawking radiation is given by 
\begin{eqnarray}
	N^{ (\mp) } (\omega,m)
	=
	\frac{ 1 }{ e^{ \beta ( \omega - m \Omega_H ) } \mp  1 },
\end{eqnarray}
where $(-)$ and $(+)$ correspond to boson and fermion, respectively.
$\beta^{-1}$ is the temperature and $\Omega_H$ is the angular velocity at the event horizon.
The explicit expressions for these quantities are given by
\begin{align}
  \beta^{-1} &= T =
  \frac{1}{4\pi R}\frac{\nu^{(N-1)/2}(1+\nu)}{(\mu+\nu)^{N/2}}\sqrt{\frac{1-\lambda}{\lambda(1+\lambda)}}\
  ,\label{eq:T_H}\\
  \Omega_H &=
  \frac{1}{R}\frac{1}{(1+\mu)^{N/2}}\sqrt{\frac{\lambda-\nu}{\lambda(1+\lambda)}}\ \label{eq:Omega_H}.
\end{align}
To avoid the ambiguity associating with the superradiance,
let us focus on the fermion case as done in~\cite{Iso:2006xj}. Including the
contribution from the antiparticles, the angular-momentum and energy
flows carried by the Hawking radiation are estimated as
\begin{eqnarray}
	&&
	J^r_{  \;\; \mathrm{ (thermal) } }
	=
	\int_0^{\infty}
	\frac{ d\omega }{ 2\pi }
	m
	\left[
		N^{(+)}( \omega, m ) - N^{(+)}( \omega, -m )
	\right]
	=
	\frac{ m^2 }{ 2\pi } \Omega_H,
	\nonumber
	\\
	&&
	T^{r}_{\;\;t \; \mathrm{ (thermal) }}
	=
	\int_0^\infty
	\frac{ d\omega }{ 2\pi }
	\omega
	\left[
		N^{(+)} (\omega, m) + N^{(+)}(\omega, -m)
	\right]
	=
	\frac{ \pi }{ 12 \beta^2 }
	+
	\frac{ m^2 }{ 4\pi } \Omega_H^2.
\end{eqnarray}
Our purpose is to derive these quantities from the viewpoint of the
anomaly cancellation. 

%%%-----------------------------------------------------------------%%%
%%%-----------------------------------------------------------------%%%
\section{ Quantum radiation eliminating the anomalies }
\label{sec:radiation}
%%%-----------------------------------------------------------------%%%
%%%-----------------------------------------------------------------%%%

%Based on the observation and definition of the effective two dimensional
%theory in the previous section, 

In this section, we will calculate angular-momentum and
energy fluxes radiated by the black rings,
by demanding the radiation eliminates the possible gauge and
gravitational anomalies on the horizon. The analysis is
basically parallel to those in Refs.~\cite{Iso:2006ut,Iso:2006xj}.

%As mentionend in the intrdocution, if one ignores the classically
%irrelevant ingoing modes near the horizon, the effective theory near
%the horizon becomes chiral, resulting in the occurance of the gauge and
%gravitational anomalyies.
%These anomalies can be eliminated by a thermal compensating radiation,
%i.e., the Hawking radiation.
%In this section, demanding the cancelation of anomalies, we calculate
%the flux of the Hawking radiation.

%%%-----------------------------------------------------------------%%%
\subsection{ Quantum field near the horizon }
\label{subsec:near-horizon}
%%%-----------------------------------------------------------------%%%

First, we investigate the behavior of a massless (real) scalar field near
the horizon of the black rings~(\ref{eq:ring}). The determinant of the
metric~(\ref{eq:ring}) and the inverse metric are
\begin{eqnarray}
\sqrt{-g} = \frac{R^4 F(x) \left[ H(x) H^2(y) \right]^{N/3}}{(x-y)^4},
\end{eqnarray}
and
\begin{multline}
  g^{\mu\nu}\partial_\mu \partial_\nu = -\frac{F(x)}{F(y)}
  \left[ \frac{H(y)}{H(x)} \right]^{N/3} \partial_t^2 \\
  + \frac{ (x-y)^2 }{ R^2 F(x) \left[ H(x) H^2(y) \right]^{N/3} }
  \left[-\frac{C^2R^2(1+y)^2 H^N(y) }{F(y)G(y)}\left(\partial_t +
  \frac{F(y)}{CR(1+y)}\partial_\psi\right)^2 \right.\\
  \left. -G(y)\partial_y^2 + G(x)\partial_x^2 +
  \frac{F(x)H^N(x)}{G(x)}\partial_\phi^2
  \right].
\end{multline}
%\begin{eqnarray}
%	&&
%	g^{ tt }
%	=
%	-
%	\frac{
%	\left[	
%		C^2 ( x-y )^2 ( 1+y )^2 + F^2(x) G(y)
%	\right]
%	H^{ N/3 }(y)
%	}
%	{
%		F(x) F(y) G(y) H^{ N/3 }(x)
%	},
%	\;\;\;\;\;\;
%	g^{ t\psi }
%	=
%	-	
%	\frac{
%		C (x-y)^2 ( 1+y ) H^{ N/3 }(y)
%	}
%	{
%		R F(x) G(y) H^{ N/3 }(x)
%	},
%	\cr
%	&&
%	g^{ \psi \psi }
%	=
%	-
%	\frac{
%		( x-y )^2 F(y)	H^{ N/3 }(y)
%	}
%	{
%		R^2 F(x) G(y) H^{ N/3 }(x)
%	},
%	\;\;\;\;\;\;
%	g^{ yy }
%	=
%	- \frac{ (x-y)^2 G(y) }{ R^2 F(x) \left[ H(x) H^2(y) \right]^{ N/3 } },
%	\cr
%	&&
%	g^{ xx }
%	=
%	\frac{ (x-y)^2 G(x) }{ R^2 F(x) \left[ H(x) H^2(y) \right]^{ N/3 } },
%	\;\;\;\;\;\;
%	g^{ \phi\phi }
%	=
%	\frac{ ( x-y )^2 H(x) }{ R^2 G(x) \left[ H(x) H^2(y) \right]^{ N/3 } }.
%\end{eqnarray}

Using these quantities, the action of the 5-dimensional scalar field is
written as
\begin{equation}
\begin{split}
	S=&\, -\frac{ 1 }{ 2 } \int d^5 \! x \sqrt{ -g }\;
	g^{\mu\nu} \partial_\mu \varphi \partial_\nu \varphi \\
	=&
	\, -\frac{ 1 }{ 2 }\int d^5 \! x 
	\frac{R^4 F(x) \left[ H(x) H^2(y) \right]^{N/3}}{(x-y)^4}
 	\bigg\{ -\frac{F(x)}{F(y)}
  	\left[ \frac{H(y)}{H(x)} \right]^{N/3} (\partial_t \varphi)^2 \\
  	&\,+ \frac{ (x-y)^2 }{ R^2 F(x)\left[ H(x) H^2(y) \right]^{N/3}}
  	\bigg[-\frac{C^2R^2(1+y)^2H^N(y)}{F(y)G(y)}\left(\partial_t \varphi +
  	\frac{F(y)}{CR(1+y)}\partial_\psi \varphi\right)^2 \\
  	&\, -G(y)(\partial_y \varphi)^2 + G(x)(\partial_x \varphi)^2 +
  	\frac{F(x)H^N(x)}{G(x)}(\partial_\phi \varphi)^2\bigg]\bigg\}.
	\label{eq:original-action}
\end{split}
\end{equation}

%\begin{eqnarray}
%	S
%	&=&
%	\frac{ 1 }{ 2 }
%	\int
%	d^5 \! x
%	\sqrt{ -g }\;
%	\varphi
%	\dalm
%	\varphi,
%	\nonumber
%	\\
%	&=&
%	\frac{ R^2 }{ 2 }
%	\int
%	d^5 \! x\;
%	\varphi
%	\Bigg\{
%		-
%		\frac{ R^2 \left[ C^2 (x-y)^2 (1+y)^2 + F^2(x) G(y) \right]
%		H^{ (N-1)/2 }(x) H^{ N }(y) }{ (x-y)^4 F(y) G(y) }
%		\partial_t^2 
%	\nonumber
%	\\
%	&&
%%	\hspace{0.2cm}
%		+
%		\frac{ 2 C R ( 1+y ) H^{ (N-1)/2 }(x) H^{ N }(y) }{ ( x-y )^2 G(y)  }
%		\partial_t \partial_\psi 
%		-
%		\frac{ F(y) H^{ (N-1)/2 }(x) H^{ N }(y) }{ (x-y)^2 G(y) }
%		\partial_\psi^2 
%	\nonumber
%	\\
%	&&
%%	\hspace{0.2cm}	
%		-
%		\partial_y
%		\left[
%			\frac{ G(y) H^{ (N-1)/2 }(x) }{ (x-y)^2 } \partial_y 
%		\right]
%		-
%		\partial_x
%		\left[
%		\frac{ G(x) H^{ (N-1)/2 }(x) }{ (x-y)^2 } \partial_x 
%		\right]
%		+
%		\frac{ F(x) H^{ (N+1)/2 }(x) }{ (x-y)^2 G(x) } \partial_\phi^2 
%	\Bigg\}\varphi.
%	\nonumber
%	\\
%	\label{eq:original-action}
%\end{eqnarray}
Taking the near-horizon limit $ y\to y_h := -1/\nu$ and leaving
dominant terms, this action reduces to
\begin{equation}
 \begin{split}
   S =&\, -\frac{ 1 }{ 2 }\int d^5 \! \tilde{x}
   \; \frac{\Delta\psi}{2\pi}
   \frac{\Delta\phi}{2\pi} \frac{R^2}{(x-y_h)^2}\\
  &\,\times
  \bigg[-\frac{C^2R^2(1+y_h)^2 H^N(y_h)}{F(y_h)G(y)}
  \left(\partial_t \varphi +
  \frac{2\pi F(y)}{CR(1+y)\Delta\psi}\partial_{\tilde{\psi}} 
\varphi\right)^2 -G(y)(\partial_y \varphi)^2 \bigg],
 \end{split}
\label{eq:actionNearHorizon-1}
\end{equation}
%\begin{eqnarray}
%	&&
%	S
%	=
%	\frac{ R^2 }{ 2 }
%	\int d^5 \! x \; \varphi
%	\Bigg\{
%		-
%		\frac{ C^2 R^2 ( 1 + y )^2 H^{ (N-1) /2 } (x) H^N (y) }{ ( x - y )^2 F(y) G(y) }
%	\times
%	\nonumber
%	\\
%	&&	
%	\times
%		\Bigg[
%			\partial_t^2 
%			+
%			\frac{ 4 \pi F(y) }{ C R ( 1 + y ) \Delta \psi }
%			\partial_t \partial_{ \tilde{ \psi } } 
%			+
%			\frac{ 4\pi^2 F^2( y ) }{ C^2 R^2 ( 1 + y )^2 ( \Delta \psi )^2 }
%			\partial_{ \tilde{ \psi } }^{2} \
%		\Bigg]
%%	\nonumber
%%	\\
%%	&&
%%	\hspace{11cm}
%	-
%	\partial_y
%	\left[
%	\frac{ G(y) H^{ (N-1)/2 }(x) }{ ( x-y )^2 }
%	\right]
%	\partial_y 
%	\Bigg\} \varphi,
%	\nonumber
%	\\
%	\label{eq:actionNearHorizon-1}
%\end{eqnarray}
where we introduce new angular coordinates 
$ \tilde{ \psi } := (2\pi/\Delta \psi) \psi $ and  
$ \tilde{ \phi } := (2\pi/\Delta \phi) \phi $
 so that $ \Delta \tilde{\psi} = 2\pi $ and 
$ \Delta \tilde{\phi} = 2\pi $ from
Eq.~(\ref{eq:deficit}).
Let us decompose the field as 
\begin{equation}
 \varphi = \sum_{m,n,l} \varphi_{mnl}(t,y)\,
  e^{im\tilde{\psi}+in\tilde{\phi}} X_l(x)\ ,
\end{equation}
where $ m,n $ $(= 0, \pm 1, \pm 2, \ldots)$ are the axial
quantum numbers. $X_l(x)$ constitutes a complete set of functions, satisfying the orthonormal relation
\begin{equation}
 \int^1_{-1} \frac{dx}{x-y_h}X_l X_{l'} = \delta_{ll'}\ .
\end{equation}
Then, we can carry out the integration with respect to $(\tilde{\psi},\tilde{\phi},x)$ and the action becomes
\begin{equation}
	S = -\frac{CR^3(1+y_h)H(y_h)^{N/2}\Delta\psi\Delta\phi}{2\sqrt{-F(y_h)}}
% 	\sum_{m,n,l}
 	\int dtdy
  	\left[ -\frac{1}{f(y)} \Big| \left[ \partial_t-imA_t(y) \right] \varphi_{mnl} \Big|^2
  	+
  	f(y) \Big| \partial_y\varphi_{mnl} \Big|^2\right],
\label{eq:actionNearHorizon}
\end{equation}
%\begin{eqnarray}
%	&&
%	S
%	=
%	\frac{ R^2 }{ 2 }
%	\int d^5 \! x \; \varphi_m^\ast
%	\Bigg\{
%		- \frac{ C R ( 1+y ) H^{ (N-1)/2 }(x) H^{ N/2 } (y) }{ ( x-y )^2 \surd{ -F(y) } }
%	\times
%	\nonumber
%	\\
%	&&
%	\hspace{6cm}
%	\left[
%		- \frac{ 1 }{ f(y) }
%		\left[
%			\partial_t - i m A_t (y)
%		\right]^2
%	+
%	\partial_y
%		f(y) \partial_y 
%	\right]
%	\Bigg\} \varphi_m,
%	\nonumber
%	\\
%	\label{eq:actionNearHorizon}
%\end{eqnarray}
where
\begin{eqnarray}
	f(y)
	&:=&
	\frac{ \sqrt{ -F(y_h) } \; G(y) }{ C R (1+y_h) H^{ N/2 } (y_h)  },
	\nonumber
	\\
	A_t (y)
	&:=&
	-
	\frac{ 2\pi F(y) }{ C R (1+y) \Delta \psi }.
	\label{eq:f-ring}
\end{eqnarray}

From Eq.~(\ref{eq:actionNearHorizon}), we see that the action 
for each mode, labeled by $m,n,l$, near the horizon is essentially
identical to that of complex scalar field in the $(1+1)$-dimensional
spacetime in the presence of a $U(1)$-gauge field.
%~
%
%\footnote{
%A few remarks may be needed for the interpretation of
%Eq.~(\ref{eq:actionNearHorizon}).
%One remark is on the $x$-dependence of the factor of the large square
%bracket in Eq.~(\ref{eq:actionNearHorizon}).
%The coordinate $x$ is a polar coordinate on the $S^2$ (roughly 
%$ x \sim \cos \theta $, see also Fig.~\ref{fg:ring-coord}). This directional
%dependence is, however, an apparent one since at the level of the
%equation of motion, the factor is clearly dropped out due to its
%overallness. This feature seems to originate from a universal nature of
%the event horizon such as the zero-th low of the black hole
%thermodynamics, and crucial for our analysis. 
%The other is on the possibility that the above factor is interpreted as
%a dilaton in the ($t,y$) sector (after fixing or suitably integrating
%out $x$). It is known, however, that such a dilaton plays no role in the
%calculation of the Hawking radiation via the anomalies at least for
%static spacetimes as in the present case~\cite{Iso:2006ut}. Therefore,
%we do not employ the interpretation for simplicity.}.
%
The charge of the
complex scalar field associated with the gauge field is $m$. The
effective 2-dimensional metric $ g_{\mu\nu} $ and gauge potential $ A $
are given by
\begin{eqnarray}
	ds^2
	&=&
	-f(r) dt^2 + f^{-1}(r) dr^2,
	\nonumber
	\\
	A
	&=&
	A_t (r) dt.
	\label{eq:effective-st}
\end{eqnarray}
Hereafter, let us regard $y$ as a``radial'' coordinate and denote $ y $
and $ y_h $ by $ r $ and $ r_H $, respectively. In the
spacetime~(\ref{eq:effective-st}), the even horizon is located at 
$ r = r_H $ where $ f( r_H ) = 0 $ (corresponding to $ G(y_h)=0 $).
From the 2-dimensional viewpoint,
the surface gravity and temperature are given by $ \kappa = f'(r_H)/2 $
and $T = \kappa/2\pi$, respectively.
The explicitly value of $T$ is given by
\begin{eqnarray}
	T
	=
	\frac{ \kappa }{ 2\pi }
	=
	\left.
	\frac{ 1 }{ 4\pi }
	\frac{ d f (r) }{ d r }
	\right|_{ r=r_H }
	=
	\frac{ 1 }{ 4 \pi R }
	\frac{ \nu^{ (N-1)/2 } ( 1+\nu ) }{ ( \mu+\nu )^{ N/2 } }
	\sqrt{
		\frac{ 1-\lambda }{ \lambda ( 1+\lambda ) }
	}.
	\label{eq:temperature}
\end{eqnarray}
One can see that this temperature coincides with Eq.~(\ref{eq:T_H}).

%%%-----------------------------------------------------------------%%%
\subsection{ Angular-momentum flux }
\label{subsec:gauge-anomaly}
%%%-----------------------------------------------------------------%%%

Let us consider the gauge anomaly near the horizon. Since the effective $ U(1) $-gauge field originates from the rotation along $ \partial_\psi $ in the original spacetime~(\ref{eq:ring}), we will see that the compensating flux against the gauge anomaly is the one of angular momentum.
Let us divide the 2-dimensional spacetime into two regions: one is a near-horizon region, $ r_H \leq r \leq r_H + \epsilon $, and the other is an outer region, $ r \geq r_H + \epsilon $, where $ \epsilon$ is a small positive parameter. If we neglect the ingoing modes in the near-horizon region, the $ U(1) $-current $ J^\mu $ obeys the following anomalous equation there,
\begin{eqnarray}
	\nabla_\mu J^{ \mu }
	=
	-\frac{ m^2 }{ 4\pi \surd{ -g } }
	\epsilon^{ \mu \nu }
	\partial_\mu A_\nu, 
\end{eqnarray}
where the notation of $ \epsilon^{ 01 } = +1 $ is used.
Assuming the stationarity of the current and that the current in the outer region is conserved, the equations for the gauge current in the outer and near-horizon regions are given by
\begin{eqnarray}	
	&&
	\partial_r J^r_{(O)} = 0,
	\nonumber
	\\
	&&
	\partial_r J^r_{(H)}
	=
	\frac{ m^2 }{ 4\pi } \partial_r A_t,
\end{eqnarray}
respectively. These equations can be integrated to give
\begin{eqnarray}
	&&
	J^r_{(O)}
	=
	c_O,
	\nonumber
	\\
	&&
	J^r_{(H)}
	=
	c_H + \frac{ m^2 }{ 4\pi }
	\left[
		A_t (r) - A_t (r_H)
	\right],
\end{eqnarray}
where $ c_O $ and $ c_H $ are integration constants.
Now, we consider the effective action $ W $ neglecting the ingoing modes at the horizon.
The variation of the action under the gauge transformation, parameterized by $ \lambda $, is calculated as
\begin{eqnarray}
	- \delta_\lambda W
	&=&
	\int d^2 \! x \sqrt{ -g } \; \lambda
	\nabla_{\mu}
	\left[
		J^\mu_{(H)} \Sigma_H (r) + J^\mu_{(O)} \Sigma_O (r)
	\right]
	\nonumber
	\\
	&=&
	\int d^2 \! x \sqrt{ -g } \; \lambda
	\left[
		\partial_r
		\left(
			\frac{ m^2 }{ 4\pi } A_t \Sigma_H
		\right)
		+
		\left(
			J^r_{ (O) } - J^r_{ (H) } + \frac{ m^2 }{ 4\pi } A_t
		\right)
		\delta( r - r_H -\epsilon )
	\right].
	\label{eq:delta W}
\end{eqnarray}
Here, $ \Sigma_O (r) $ and $ \Sigma_H (r) $ are the supports of $ J^\mu_{(O)} $ and $ J^\mu_{(H)} $, respectively, defined by the step function $ \Theta $ as
\begin{eqnarray}
	\Sigma_O (r) := \Theta ( r -r_H -\epsilon ),
	\;\;\;\;\;
	\Sigma_H(r) := 1 - \Theta ( r -r_H -\epsilon ).
\end{eqnarray}
The first term in Eq.~(\ref{eq:delta W}) should be canceled by the quantum effect of the classically irrelevant ingoing modes. On the other hand, the coefficient of the delta function should vanish for the anomaly at the horizon to be canceled, in other words, for the gauge invariance to persist at the quantum level. This requirement reads  
\begin{eqnarray}
	c_O = c_H - \frac{ m^2 }{ 4\pi } A_t(r_H) .
\end{eqnarray}
To fix the values of the coefficients, we have to impose a boundary condition.
The boundary condition that we adopt is that the vanishing of the covariant current~\cite{Bardeen:1984pm} at the event horizon~\cite{Iso:2006wa}~\footnote{
See Refs.~\cite{Iso:2006wa,Murata:2006pt} for the discussion on the relation between this kind of boundary condition and the choice of the vacuum state.
}. The covariant current, denoted by $\tilde{J}^\mu$, is given by
\begin{eqnarray}
	\tilde{J}^{\mu}
	=
	J^{\mu}
	-
	\frac{ m^2 }{ 4\pi \surd{-g} }
	A_\lambda \epsilon^{ \lambda \mu },
\end{eqnarray}
and satisfies
\begin{eqnarray}
	\nabla_\mu \tilde{ J }^\mu
	=
	\frac{ m^2 }{ 4\pi \surd{-g} }
	\epsilon_{\mu\nu} F^{ \mu\nu },
\end{eqnarray}
where $ F_{\mu\nu} := 2 \; \partial_{ \;[\mu } A_{ \nu ] } $ is the field strength of the gauge field.
Since $ \tilde{J}^r = J^r + ( m^2 / 4\pi) A_t (r) \Sigma_H (r) $ in the present case,
the boundary condition leads to
\begin{eqnarray}
	c_O
	=
	- \frac{ m^2 }{ 2\pi } A_t (r_H)
	=
	\frac{ m^2 }{ 2 \pi  R ( 1+\mu )^{ N/2 } }
	\sqrt{ \frac{ \lambda - \nu  }{ \lambda ( 1+\lambda ) } },
	\label{eq:cO}
\end{eqnarray}
where Eq.~(\ref{eq:deficit}) is used. 
This is the gauge current (therefore, angular-momentum flux) in the
outside region, obtained by imposing the cancellation of the gauge
anomaly at the horizon. This value exactly coincides with the
angular-momentum flux derived by the
Planckian distribution in Sec.~\ref{sec:fluxes}.

%%%-----------------------------------------------------------------%%%
\subsection{ Energy flux }
\label{subsec:gravitational-anomaly}
%%%-----------------------------------------------------------------%%%

Now, we calculate the energy flux by imposing
the vanishing of anomalies at the horizon.
Due to the existence of the gauge current, the energy momentum of the 2-dimensional theory is not conserved even classically. The appropriate Ward-Takahashi identity with the gravitational anomaly, $ \mathcal{A}_{\nu} $, added is given by~\cite{Iso:2006xj}, 
\begin{eqnarray}
	\nabla_\mu T^{\mu}_{\;\;\nu}
	=
	F_{ \mu\nu } J^{\mu}
	+
	A_\nu \nabla_\mu J^{\mu}
	+
	\mathcal{A}_\nu.
\end{eqnarray}
The consistent current~\cite{Alvarez-Gaume:1983ig,Bertlmann:2000da} of the gravitational anomaly is given by
\begin{eqnarray}
	\mathcal{A}_\nu
	=
	- \frac{ 1 }{ 96\pi \surd{ -g } }
	\epsilon^{ \beta \delta }
	\partial_\delta \partial_\alpha \Gamma^{ \alpha }_{ \;\; \nu \beta }
	=:
	\frac{ 1 }{ \surd{-g} }
	\partial_\alpha N^{ \alpha }_{ \;\;\nu }.
	\label{eq:anomaly}
\end{eqnarray}
The components of $ { N^\alpha }_\nu $ are
\begin{eqnarray}
 {N^t}_t={N^r}_r=0,\;\;
 {N^r}_t=-\frac{1}{192\pi}(f^{\prime 2}+f''f), \;\;
 {N^t}_r= \frac{1}{192\pi f^2}(f^{\prime 2}-f''f) .
\end{eqnarray}
Assuming the absence and presence of anomalies in the outer and near-horizon regions, respectively, the Ward-Takahashi identities become
\begin{eqnarray}
	&&
	\partial_r T_{ (O)~t }^{ ~~~r }
	=
	F_{rt} J^{r}_{(O)},
	\nonumber
	\\
	&&
	\partial_r T_{ (H)~t }^{ ~~~r }
	=
	F_{rt} J^{r}_{(H)}
	+
	A_t \partial_r J^{r}_{(H)}
	+
	\partial_r N^{r}_{\;\;t}.
\end{eqnarray}
These equations can be integrated to give
\begin{eqnarray}
	&&
	T_{ (O)~t }^{ ~~~r }
	=
	a_O
	+
	c_O A_t (r),
	\nonumber
	\\
	&&
	T_{ (H)~t }^{ ~~~r }
	=
	a_H
	+
	\int^r_{r_H} dr \; \partial_r
	\left(
		c_O A_t + \frac{ m^2 }{ 4\pi } A_t^2 + N^{r}_{\;\;t}
	\right),
\end{eqnarray}
where $ a_O $ and $ a_H $ are integration constants.
By an infinitesimal coordinate transformation in the time direction, parameterized by $\xi^t$, the effective action changes as
\begin{eqnarray}
	- \delta_\xi W
	&=&
	\int d^2 \! x
	\sqrt{ -g } \;
	\xi^t \nabla_\mu
	\left[
		T_{ (H) ~t }^{ ~~~\mu } \Sigma_H (r)
		+
		T_{ (O) ~t }^{ ~~~\mu } \Sigma_O (r)
	\right]
	\nonumber
	\\
	&=&
	\int d^2 \!x \;
	\xi^{t}
	\Bigg[
		c_O \partial_r A_t
		+
		\partial_r
		\left\{\left(
		\frac{ m^2 }{ 4\pi } A_t^2
		+
		N^{r}_{\;\;t}
		\right)\Sigma_H\right\}
		+
	\nonumber
	\\
	&&
	\hspace{4cm}
		\left(
			T_{ (O)~t }^{ ~~~r } - T_{ (H)~t }^{ ~~~r }
			+ \frac{ m^2 }{ 4\pi } A_t^2
			+ N^{ r }_{ \;\;t }
		\right)
		\delta ( r-r_H-\epsilon )
	\Bigg].
	\label{eq:variation1}
\end{eqnarray}
The first term is purely classical effect of the background current flow. The second term should be canceled by the classically irrelevant ingoing modes again. The coefficient of the delta function should vanish to save the diffeomorphism invariance at the quantum revel. This requirement leads to
\begin{eqnarray}
	a_O
	=
	a_H
	+
	\frac{ m^2 }{ 4\pi } A_t^2 (r_H)
	-
	N^{r}_{\;\;t} (r_H).
\end{eqnarray}
To know $ c_O $, we have to determine $ c_H $ by imposing a boundary condition on the anomalous current. We impose the vanishing of the covariant current at the horizon again since the boundary condition should be diffeomorphism invariant. In the present case, the covariant energy-momentum tensor is given
\begin{eqnarray}
	\tilde{T}^{ r }_{ \;\;t }
	=
	T^{ r }_{ \;\; t }
	+
	\frac{ 1 }{ 192 \pi }
	\left(
		f f^{ \prime \prime } - 2 f^{ \prime 2 }
	\right).
\end{eqnarray}
The vanishing of this covariant current at the horizon determine $ a_H $ as
\begin{eqnarray}
	a_H
	=\frac{f'(r_H)}{96\pi} = 
	\frac{ \kappa^2 }{ 24\pi }\ ,
\end{eqnarray}
where $\kappa = f'(r_H)/2$. 
Thus, we can determine $a_O$, the total flux of the quantum radiation, as
\begin{eqnarray}
	a_O
	=
	\frac{ \kappa^2 }{ 48 \pi }
	+
	\frac{ m^2 ( \lambda-\nu ) }{ 4 \pi R^2 ( 1+\mu )^N \lambda ( 1+\lambda ) }.
	\label{eq:aO}
\end{eqnarray}
This is the energy flux in the
outside region, obtained by imposing the cancellation of the gauge
anomaly at the horizon.
This value exactly coincides with the
energy flux derived by the
Planckian distribution in Sec.~\ref{sec:fluxes}.

%%%-----------------------------------------------------------------%%%
\subsection{ Fluxes for Myers-Perry black hole }
\label{subsec:MP}
%%%-----------------------------------------------------------------%%%

We saw in Sec.~\ref{subsec:dipole-ring} that the metric~(\ref{eq:ring}) describes the Myers-Perry black hole in the suitable limit.
Therefore, it will be important that our results on the quantum radiation (i.e., the fluxes of angular-momentum and energy) reproduce those for the Myers-Perry black hole.
First, let us consider the limit of the angular-momentum flow~(\ref{eq:cO}) and energy flow~(\ref{eq:aO}). By applying the limiting procedure described in Sec.~\ref{subsec:dipole-ring} (i.e., $\mu \to 0$, $ \lambda, \nu \to 1 $ and $ R \to 0 $ with $M$ and $a$ kept finite) to Eqs.~(\ref{eq:cO}) and (\ref{eq:aO}), we have
\begin{eqnarray}
	&&
	c_O
	\;
	\to
	\;
	m^2 
	\frac{  a  }{ 2 \pi M },
	\nonumber
	\\
	&&
	a_O
	\;
	\to
	\;
	\frac{ M-a^2 }{ 48\pi M^2 }
	+
	m^2
	\frac{  a^2 }{ 4 \pi M^2  }.
\end{eqnarray}
We can check that these values coincide with those for the Myers-Perry black hole with a single rotation~\cite{Iso:2006xj}.
Regarding also the temperature~(\ref{eq:temperature}), we can take safely the neutral limit ($ \mu\to 0 $) and the limit to the Myers-Perry black hole successively:
\begin{eqnarray}
	T
	\;
	\to
	\;
	\frac{ 1 }{ 4\pi R } ( 1+\nu )
	\sqrt{ \frac{ 1-\lambda }{ \lambda\nu(1+\lambda) } }
	\;
	\to
	\;
	\frac{ \sqrt{ M-a^2 } }{ 2 \pi M }.
	\label{eq:temp-limits}
\end{eqnarray}
We see that the resultant expression after the first limit is the temperature of the Emparan-Reall black ring (e.g., see \cite{Emparan:2006mm} for the same parametric expression) and the expression after the second limit is the temperature of the Myers-Perry black hole~\cite{Myers:1986un}.

Thus, we can say that our analysis on the anomaly cancellation in the black rings covers the (essential parts of) analyses on the Myers-Perry black holes in Refs.~\cite{Murata:2006pt} and~\cite{Iso:2006xj}.
%
%
%
%
%%%%-----------------------------------------------------------------%%%
%%%%-----------------------------------------------------------------%%%
\section{ Conclusion }
\label{sec:conclusion}
%%%-----------------------------------------------------------------%%%
%%%-----------------------------------------------------------------%%%

We have calculated the fluxes of angular momentum and energy radiated from the 5-dimensional rotating dipole black rings with the requirement that the possible diffeomorphism anomalies at the horizons should be canceled by the radiation. We have seen that this requirement with the physically reasonable boundary conditions fixes the values of flux to strictly coincide with ones calculated by the integration of Planckian spectrum. The temperature of black rings also has been correctly predicted by the near-horizon behavior of the quantum field. Since the class of black rings considered in this paper contains the Emparan-Reall neutral black ring and the Myers-Perry black hole, the limits to these solutions have been investigated and the fluxes and temperatures for these black objects have been recovered. 

The results suggest that the effective theory of quantum fields near the horizons can be reduced to two dimensional one in a wide class of black objects even with non-trivial horizon topologies.
Various generalization of this work will be possible.
The generalization to the recently-discovered black holes in higher dimensions, which can have multiple angular momenta~\cite{Kudoh:2006xd,Pomeransky:2006bd} and/or multiple horizons~\cite{Elvang:2007rd,Iguchi:2007is,Yazadjiev:2007cd}, would be straightforward but will be important, since the knowledge of the thermal properties and Hawking radiation of black holes is essential to understand the phase structure and their evolution. The most challenging work would be to derive the thermal spectrum~\cite{Iso:2007kt} and the entropy formula in this line.

%%%-----------------------------------------------------------------%%%
%%%-----------------------------------------------------------------%%%

%%%-----------------------------------------------------------------%%%
%%%-----------------------------------------------------------------%%%
\medskip
\section*{Acknowledgements}
UM would like to thank Masato Nozawa for informative conversations and critical reading of the manuscript.
This work is supported in part by a Grant for The 21st Century COE Program (Holistic Research and Education Center for Physics Self-Organization Systems) at Waseda University.

\medskip
\noindent
\textit{Note added.---}
After completing the analysis,
we found an independent work by Bin Chen and Wei He~\cite{Chen:2007pp} on the same subject. 
%%%-----------------------------------------------------------------%%%
%%%-----------------------------------------------------------------%%%

%%%%%%%%%%%%%%%%%%%%%%%%%%%%%%%%%%%%%%%%%%%%%%%%%%%%%%%%%%%%%%%%%%%%%%%%%%%%%%%

%%%%%%%%%%%%%%%%%%%%%%%%%%%%%%%%%%%%%%%%%%%%%%%%%%%%%%%%%%%%%%%%%%%%%%%%%%%%%%%
%\bibliographystyle{unsrt} 
%\bibliography{GAnomaly.bib}
%%%%%%%%%%%%%%%%%%%%%%%%%%%%%%%%%%%%%%%%%%%%%%%%%%%%%%%%%%%%%%%%%%%%%%%%%%%%%%%

%%%%%%%%%%%%%%%%%%%%%%%%%%%%%%%%%%%%%%%%%%%%%%%%%%%%%%%%%%%%%%%%%%%%%%%%%%%%%%%
%\bibliographystyle{apsrev} 
\bibliographystyle{unsrt} %% plain, unsrt, alpha, abbrv, acm, apalike
%% \bibliography{BH02.bib}

%%%%%%%%%%%%%%%%%%%%%%%%%%%%%%%%%%%%%%%%%%%%%%%%%%%%%%%%%%%%%%%%%%%%%%%%%%%%%%%
%\input{g-anomaly-2007-05-22B_bbl.tex}

\end{document}